\documentclass[10pt,conference,letterpaper]{IEEEtran}

\ifCLASSINFOpdf
  \usepackage[pdftex]{graphicx}
\else
  \usepackage[dvips]{graphicx}
\fi
\usepackage[paper=letterpaper,left=0.65in,right=0.65in,top=0.7in,bottom=1.0in]{geometry}

\usepackage{wrapfig}

\usepackage{enumitem}
\usepackage{esvect}
\setlist[enumerate]{leftmargin= 0.5 cm}
\setlist[itemize]{leftmargin=0.3 cm}

\usepackage{epstopdf}
\usepackage{graphicx}
\usepackage[dvipsnames]{xcolor}
\definecolor[named]{ACMBlue}{cmyk}{1,0.1,0,0.1}
\definecolor[named]{ACMYellow}{cmyk}{0,0.16,1,0}
\definecolor[named]{ACMOrange}{cmyk}{0,0.42,1,0.01}
\definecolor[named]{ACMRed}{cmyk}{0,0.90,0.86,0}
\definecolor[named]{ACMLightBlue}{cmyk}{0.49,0.01,0,0}
\definecolor[named]{ACMGreen}{cmyk}{0.20,0,1,0.19}
\definecolor[named]{ACMPurple}{cmyk}{0.55,1,0,0.15}
\definecolor[named]{ACMDarkBlue}{cmyk}{1,0.58,0,0.21}
\usepackage{hyperref}
\hypersetup{colorlinks,
      linkcolor=blue,
      citecolor=ACMRed,
      urlcolor=ACMBlue,
      filecolor=ACMDarkBlue}

\usepackage{subcaption} 
\usepackage{amsmath}
\usepackage{amsbsy}
\usepackage{diagbox}
\usepackage{array}
\newcolumntype{y}[1]{>{\centering\let\newline\\\arraybackslash\hspace{0pt}}p{#1}}

\allowdisplaybreaks

\usepackage{amsthm}
\usepackage[ruled, lined, linesnumbered, commentsnumbered, longend]{algorithm2e}
\usepackage[noend]{algpseudocode}

\usepackage{amssymb}
\usepackage[mathscr]{eucal}
\usepackage{url}
\usepackage{thmtools}
\usepackage{mathtools}
\usepackage{color}
\usepackage{multirow}
\usepackage{soul} 
\usepackage{siunitx}
\usepackage{comment}
\usepackage{cite}

\usepackage[utf8]{inputenc}
\usepackage{booktabs}

\usepackage{diagbox}

\theoremstyle{definition}

\declaretheoremstyle[
  headfont=\normalfont\bfseries,
  numbered=unless unique,
  bodyfont=\normalfont,
  qed={$\blacksquare$}
]{exmpstyle2}

\theoremstyle{definition}

\DeclareMathAlphabet{\mathpzc}{OT1}{pzc}{m}{it} 

\usepackage{acro}

\acsetup{format/first-long =\itshape} 

\newcommand{\Title}{\huge{{\systemNameAbbr}: An End-to-End Resource-Aware Scheduler for Machine Learning Application Requests}}

\DeclareAcronym{ML}{
short = ML,
long = machine learning
}
\DeclareAcronym{GPU}{
short = GPU,
long = graphics processing unit,
short-plural-form = GPUs,
long-plural-form = graphics processing units,
}
\newcommand{\systemNameAbbr}{StraightLine}
\newcommand{\layerOne}{Model containerization}
\newcommand{\LayerOne}{Model Containerization}
\newcommand{\layerTwo}{Container customization}
\newcommand{\LayerTwo}{Container Customization}
\newcommand{\layerThr}{Real-time resource-aware scheduling}
\newcommand{\LayerThr}{Real-time Resource-aware Scheduling}
\newcommand{\layerOneSmall}{model development abstraction}
\newcommand{\layerTwoSmall}{multiple implementation deployment}
\newcommand{\layerThrSmall}{real-time resource placement}

\definecolor{dv}{RGB}{148, 0, 211}

\renewcommand{\baselinestretch}{0.99}
\definecolor{orange}{rgb}{1,0.5,0}
\graphicspath{ {graph/} {img/}}

\IEEEoverridecommandlockouts
\def\BibTeX{{\rm B\kern-.05em{\sc i\kern-.025em b}\kern-.08em T\
kern-.1667em\lower.7ex\hbox{E}\kern-.125emX}}

\begin{document}

\onecolumn

\title{\Title\vspace{0.1cm}}

\author{
Cheng-Wei Ching\IEEEauthorrefmark{1},
Boyuan Guan\IEEEauthorrefmark{2},
Hailu Xu\IEEEauthorrefmark{3},
Liting Hu\IEEEauthorrefmark{1} \vspace{.5cm}\\
\IEEEauthorrefmark{1}Department of Computer Science and Engineering, University of California Santa Cruz, Santa Cruz, CA, USA\\
\IEEEauthorrefmark{2}Geographic Information System (GIS) Center of Florida International University, Miami, FL, USA\\
\IEEEauthorrefmark{3}Department of Computer Engineering and Computer Science, California State University, Long Beach, CA, USA \vspace{0.1cm}\\
Corresponding author's email: lhu82@ucsc.edu (Liting Hu)
}

\maketitle

\begin{abstract}
The life cycle of machine learning (ML) applications consists of two stages: model development and model deployment. However, traditional ML systems (e.g., training-specific or inference-specific systems) focus on one particular stage or phase of the life cycle of ML applications. These systems often aim at optimizing model training or accelerating model inference, and they frequently assume homogeneous infrastructure, which may not always reflect real-world scenarios that include cloud data centers, local servers, containers, and serverless platforms.
We present StraightLine, an end-to-end resource-aware scheduler that schedules the optimal resources (e.g., container, virtual machine, or serverless) for different ML application requests in a hybrid infrastructure. The key innovation is an empirical dynamic placing algorithm that intelligently places requests based on their unique characteristics (e.g., request frequency, input data size, and data distribution). In contrast to existing ML systems, StraightLine offers end-to-end resource-aware placement, thereby it can significantly reduce response time and failure rate for model deployment when facing different computing resources in the hybrid infrastructure.


\end{abstract}



\IEEEpeerreviewmaketitle

\section{Introduction}
\label{sec: introduction}

\Ac{ML} has recently evolved from a small field of academic research to an applied field. 
For Industry 4.0, factories are increasingly utilizing ML for tasks such as anomaly detection and edge robotics \cite{industry4.0,hao2019efficient, liang2019hdso}.
In the realm of autonomous vehicles, a plethora of ML-based applications have emerged, including object tracking, object detection, driving decision-making, and many others applications
\cite{kong2021fedvcp,posner2021federated, zhang2021distributed}.

The life cycle of \ac{ML} applications can be categorized into two stages: model development and model deployment.
In model development, \ac{ML} application developers need to go over three major phases: (i) data management: preparing data that is needed to build a \ac{ML} model; (ii) model training: model selection and training (iii) model verification: to ensure the model adheres to certain functional and performance requirements~\cite{kuo2021energy, ching2024totoro,ching2020model,ching2023dual,lan2024improved}.
In model deployment, developers need to consider two major phases: (i) infrastructure building: building the infrastructure to run the \ac{ML} model; (ii) model implementation: implementing the \ac{ML} model itself in a form that can be consumed and supported. In different phases and stages, the requirements of resources are quite diverse. For example, \acp{GPU} are indispensable to accelerate training process, whereas in model verification it is suitable to allocate lightweight \acp{GPU} or CPU to execute model inference~\cite{li2024comprehensive, li2024feature, lan2024asynchronous}.
However, traditional ML systems focus on one particular stage or phase of the life cycle of ML applications. These systems often aim at optimizing model training or accelerating model inference, and they frequently assume homogeneous infrastructure for model training or model inference, which may not always reflect real-world scenarios. In real-world industrial deployment of \ac{ML} applications, the infrastructure is often hybrid, which includes cloud data centers, local servers, containers, and serverless platforms (e.g., AWS Lambda \cite{AmazonLambda}).

In this paper, we present StraightLine, an end-to-end resource-aware scheduler that schedules the optimal resources (e.g., container, virtual machine, or serverless process) for different ML application requests in a hybrid infrastructure. The key innovation is an empirical dynamic placing algorithm that intelligently places requests based on their unique characteristics (e.g., request frequency, input data size, and data distribution). In contrast to existing ML systems, StraightLine offers end-to-end resource-aware placement, thereby it can significantly reduce response time and failure rate for model deployment when facing different computing resources in the hybrid infrastructure. We run real-world experiments in a hybrid testbed (RESTful APIs, serverless processes, Docker containers, and virtual machines) and demonstrate StraightLine's effectiveness in adapting to heterogeneous resources.

\section{Design}
\label{sec:design}

\subsection{Overview}
\label{subsec :overview}
\begin{wrapfigure}{r}{0.4\linewidth}
\center
 \includegraphics[width=1.\linewidth]{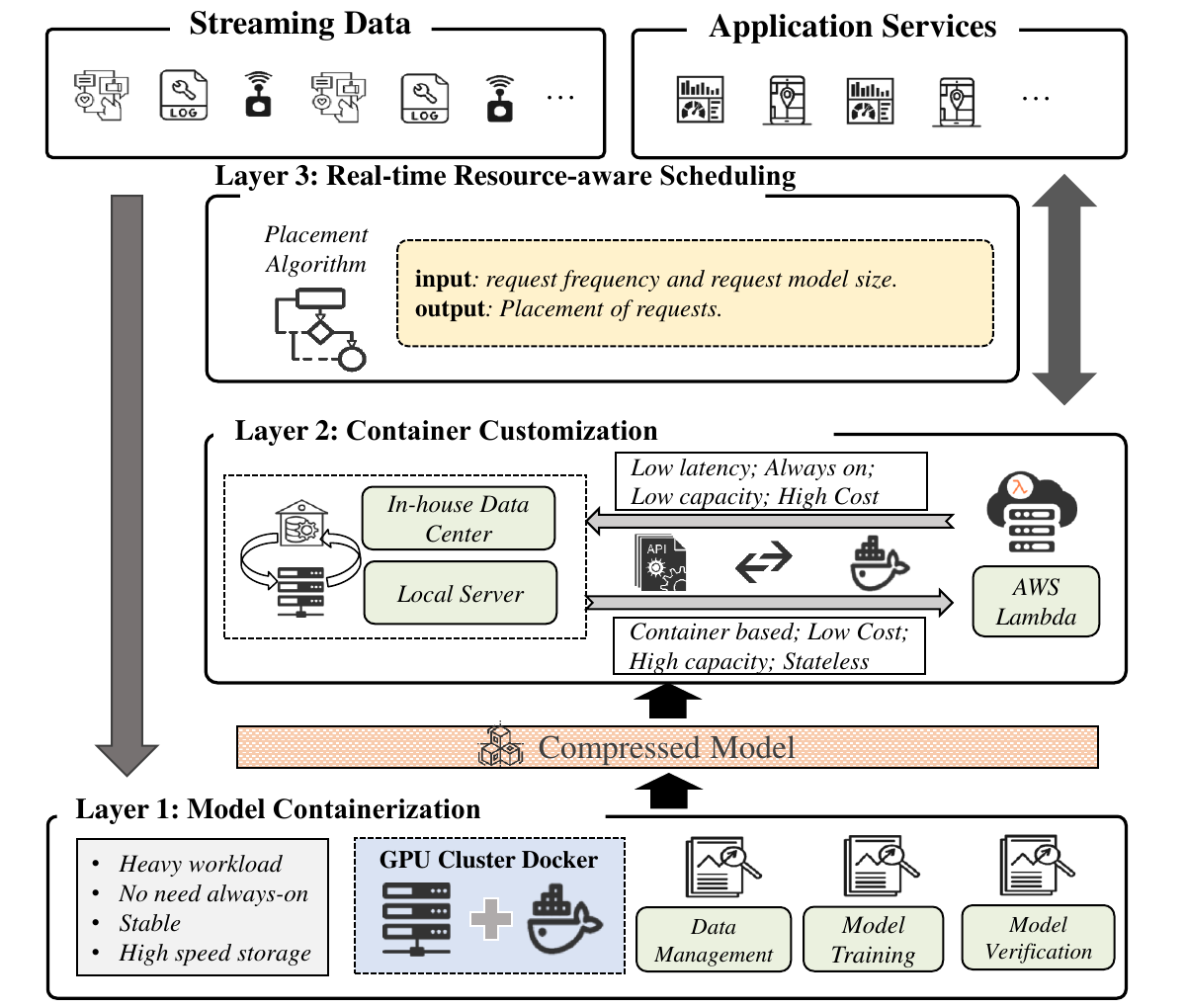}
\caption{\textit{The workflow of {\systemNameAbbr}}. 
}
\label{fig:3 layers}
 \vspace{-0.3in}
\end{wrapfigure}

\quad \enspace As shown in Figure \ref{fig:3 layers}, {\systemNameAbbr} consists of three layers: 
\textit{\layerOneSmall}, 
\textit{\layerTwoSmall}, \textit{\layerThrSmall}.

\textbf{Layer 1: {\layerOne}}. For the phases of model training, we build up powerful NVIDIA-Docker \cite{NVIDIADocker} to offer plug-and-go provisioning instead of using a traditional GPU cluster. For model verification, we build up a lightweight NVIDIA-Docker to verify the model performance. The final trained and verified \ac{ML} models are compressed and serve as the implementational fundamentals for \ac{ML} applications.

\textbf{Layer 2: {\layerTwo}}. Based on different compressed \ac{ML} models, we build up corresponding RESTful APIs, serverless applications, and Docker containers. The key innovation is to leverage Docker containers to adapt to different computing environments. 

\textbf{Layer 3: {\layerThr}}.  We design an empirical dynamic placing algorithm that intelligently places different ML application requests based on their unique characteristics (e.g., request frequency, input data size, and data distribution) in a hybrid infrastructure, with the goal of minimizing the latency and response time.

\subsection{\LayerOne}
\label{subsec: layerone}


To achieve plug-and-go and stable functionalities in model development, we containerize the phases with NVIDIA-Docker \cite{NVIDIADocker}. NVIDIA-Docker is a thin wrapper on top of Docker \cite{NVIDIADocker}. When creating a container using NVIDIA-Docker, we specify the information of the CUDA devices, volumes, and libraries in NVIDIA-Docker and it creates a container with this information. This allows us to have a plug-in GPU-aware container to implement high-performance and stable model development. Moreover, when the running tasks finish, the provisioned resources will be released back to the infrastructure automatically.

In practice, we configure more GPU resources in NVIDIA-Docker for data management and model training, whereas configuring less GPU resource for model verification since model verification is lightweight compared to model training in terms of computation complexity. After model development is completed, the trained model will be compressed as a \texttt{.h5} file and go through the stage of model deployment.

\subsection{\LayerTwo}

{\systemNameAbbr} is designed for hybrid infrastructure so compressed models are implemented in three different ways: 1) local web server, 2) RESTful APIs, or 3) serverless computing. However, it is likely that the hybrid infrastructure cannot offer a compatible environment to many heterogeneous \ac{ML} applications. 
Each computing unit in the hybrid infrastructure may run different operating systems, ML application runtime (e.g., TensorFlow \cite{abadi2016tensorflow}, PyTorch \cite{NEURIPS2019_9015}, PyWren \cite{jonas2017occupy}, etc.), and language environments (e.g., Python, Java, or R). It is imperative to consider the implementation difficulty resulting from software environment conflicts.

\begin{wrapfigure}{r}{0.3\linewidth}
 \centering
 \vspace{-0.15in}
 \includegraphics[width=1.\linewidth]{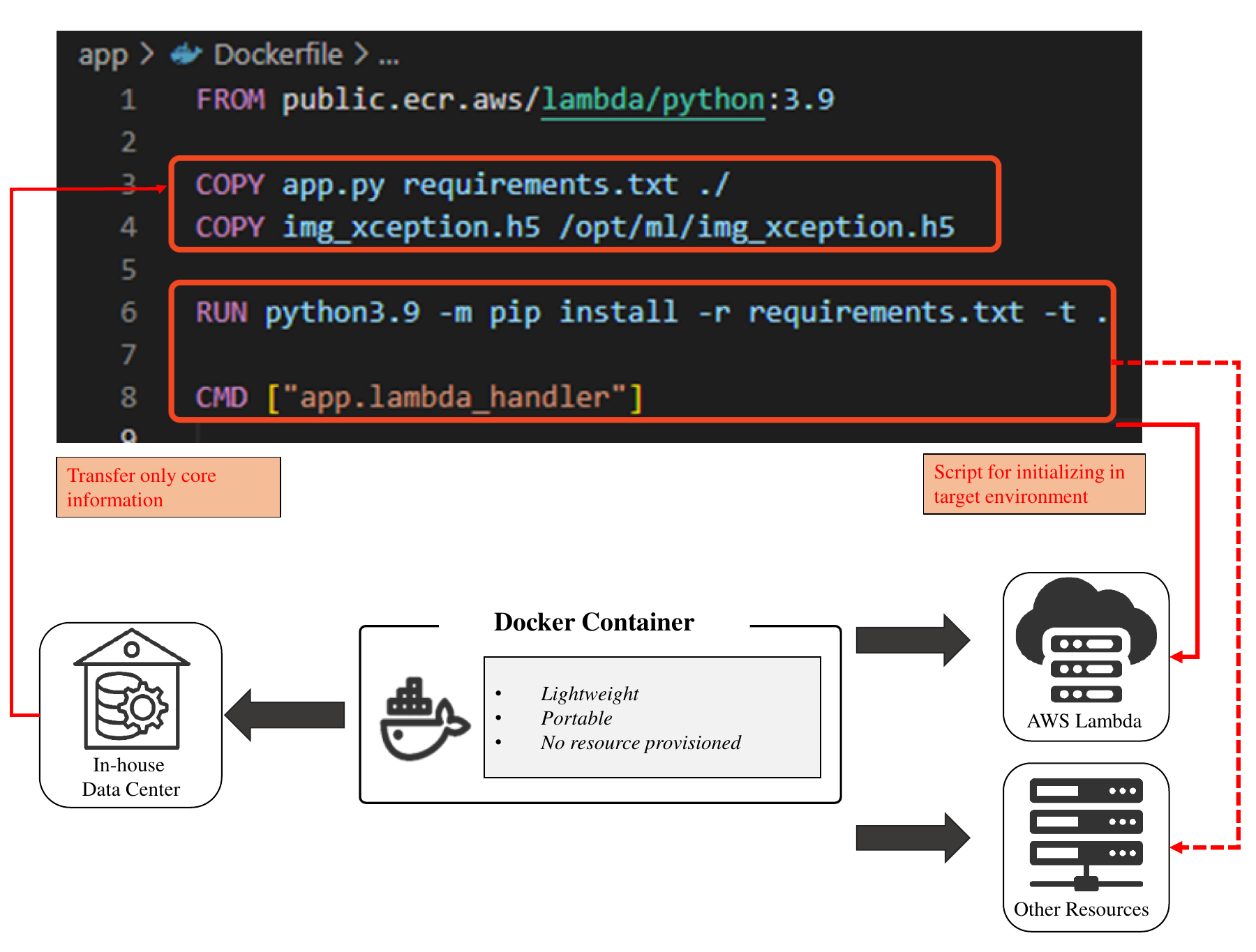}
 \caption{\emph{The core information in the dockerfile.}}
 \label{fig:webserviceDockerResponse}
 \vspace{-0.2in}
\end{wrapfigure}

We further offer the implementation of containerized \ac{ML} applications. As shown in Figure \ref{fig:webserviceDockerResponse}, a containerized \ac{ML} application only contains core information (e.g., model weights, and inference requirements) and the target environment (e.g., ML application runtime and language environment). Once a containerized \ac{ML} application is triggered in the infrastructure, it can connect to the specified target environment and resources. When the task is finished, the provisioned resources will be released back to the infrastructure. Moreover, we can execute cross-platform \ac{ML} implementation by specifying different target environments, such as different versions of Linux (e.g., Ubuntu), Windows or serverless environments.

In practice, we use the Flask \cite{flask} Python library to implement RESTful APIs for \ac{ML} implementation since most machine learning libraries are built on Python. For serverless computing, we use AMS Lambda \cite{AmazonLambda} to implement \ac{ML} applications.

\subsection{\LayerThr}

\begin{algorithm}[t]
    \caption{Empirical Dynamic Placing Algorithm}
    \label{algo:dynamicPlace}
    \KwIn{A set of requests $R$, and each request $r$'s request id $r_{id}$ request frequency at time $t$ $f_t$, request data size $r_d$, frequency threshold $F$, data size threshold $D$, available resources of Flask $S_F$, and Docker $S_D$.}
    \KwOut{The placement $K$ of the set of requests $R$.}

    $K\leftarrow \emptyset$ \;
    \For{$r \in R$}{
        \eIf{$f_t > F$ and $r_d < D$ \label{pseudo:serverless}}{
            $K\leftarrow K \cup \{(r_{id}, \texttt{serverless})\}$ \;\tcc{Run request $r$ on serverless platform}
        }{
        \eIf{$r_d > D$ \label{pseudo:docker}}{
             $K\leftarrow K \cup \{(r_{id}, \texttt{docker})\}$ \;\tcc{Run request $r$ on docker platform}
        }{
            \eIf{$S_F$ is not empty \label{pseudo:flask}}{
                $K\leftarrow K \cup \{(r_{id}, \texttt{Flask})\}$ \;
                \tcc{Run request $r$ on Flask}
            }{
                \eIf{$S_D$ is not empty \label{pseudo:docker_if_no_flask}}{
                    $K\leftarrow K \cup \{(r_{id}, \texttt{docker})\}$ \;\tcc{Run request $r$ on docker platform}
                }{\label{pseudo:serverless_if_no_flask}
                    $K\leftarrow K \cup \{(r_{id}, \texttt{serverless})\}$ \;\tcc{Run request $r$ on serverless platform}
                }
            }
        }
        }
     }
    \Return Placement $K$    
\end{algorithm}


In real-world industrial infrastructure, multiple computing resources are available. Moreover, uncertainties, such as request frequency, and request data size, arise in running \ac{ML} applications. For example, a single image classification request can be done by Flask API while a batch of 100 image classification simultaneous requests have to be sent to Amazon Lambda to handle. Therefore, the challenge is \textit{how to allocate appropriate resources based on different \ac{ML} application requests?}

We propose an empirical dynamic placing algorithm in Algorithm \ref{algo:dynamicPlace} to find the optimal resource type for an upcoming \ac{ML} application's request. 
Suppose that there is a set of requests $R$ in the waiting queue. Denote request id and request data size of request $r \in R$, request frequency at time $t$ by $r_{id}, r_d, f_t$, respectively. Also, let $F$ and $D$ denote the request frequency and request data size thresholds, respectively.
In line \ref{pseudo:serverless}, we first investigate the frequency and data size of a request. If the current request frequency is larger than the frequency threshold and data size is smaller than the data size threshold, we deploy the request on the serverless platform. 
This is because the web server may have been handling heavy workloads, and a small data size brings small communication overhead to transmit requests to serverless resources. Therefore, it is suitable to offload the request to the serverless platform. 
In line \ref{pseudo:docker}, we consider deploying the request with large data size on the Docker platform as the requests with large data sizes usually tolerate longer response time. For example, the analysis of high-resolution medical images can tolerate higher response time as medical diagnosis focuses on accurate analysis instead of quick responses. Hence, it is suitable to put the requests in the queue of Docker. 
In line \ref{pseudo:flask}, we deploy the requests using Flask (i.e., local web server) if the request has moderate data size and the current request frequency is low. The main reason is that such a request matches the characteristics of Flask: lower response time but unable to process requests with high request frequencies and large data sizes.
In lines \ref{pseudo:docker_if_no_flask} and \ref{pseudo:serverless_if_no_flask}, requests with moderate data sizes are processed by Docker and serverless platforms only under specific conditions: when Flask is unavailable for more requests and the request frequency is moderate. Within this setup, Docker is given priority. Requests allocated to a Docker container will be run using RESTful APIs.

\section{Evaluation}
\label{sec:eval}


We evaluate {\systemNameAbbr} on a real-world hybrid infrastructure testbed. We conducted preliminary experiments to show the varying failure rates, session lengths, and response times of ML models in different computational environments. This demonstrates {\systemNameAbbr}'s effectiveness in reducing response time and failure rates for model deployment when encountering diverse computing resources in a hybrid infrastructure.

\subsection{Setup}
\textbf{Real-world Testbed.} 
We set up an in-house data center using a 2-node GPU-ready in-house data center. Each node contains 36 cores of Intel(R) Core(TM) i9-7980XE CPU at 2.6Hz, 64GB of memory, and a NVIDIA GeFore GTX 1080 Ti \ac{GPU} card with 11GB of dedicated \ac{GPU} memory. NVIDIA-Docker's equipment is similar to the in-house data center but only has 36 cores CPU, 64GB memory, and 1 GPU resource. 
For model deployment, we build up a hybrid infrastructure that consists of a local web server with 12 cores of Intel(R) Xeon(R) E-2176M CPU at 2.70GHz, 32GB of memory, and serverless computing of AWS Lambda \cite{AmazonLambda} with \texttt{us-east2} and 5GM memory.

 For RESTful service, we use TensorFlow \cite{abadi2016tensorflow} as the \ac{ML} application runtime and Flask \cite{flask} to expose it as the RESTful APIs. The web server is the Internet Information Server (IIS)%
hosted on a Windows Server 2012. The WFastCgi module%
is used to bridge the Flask API and the IIS server. The Flask application is set up as a virtual site and the actual inference function is set as a route of the virtual site. 
For the network, we set the \texttt{connectionTiemout} as 5 minutes, \texttt{maxBandwidth} and \texttt{maxConnections} as 4GB, and \texttt{maxURLSegments} with 32 pieces.

\begin{wrapfigure}{l}{0.5\linewidth}
 \centering

\includegraphics[width=.8\linewidth]{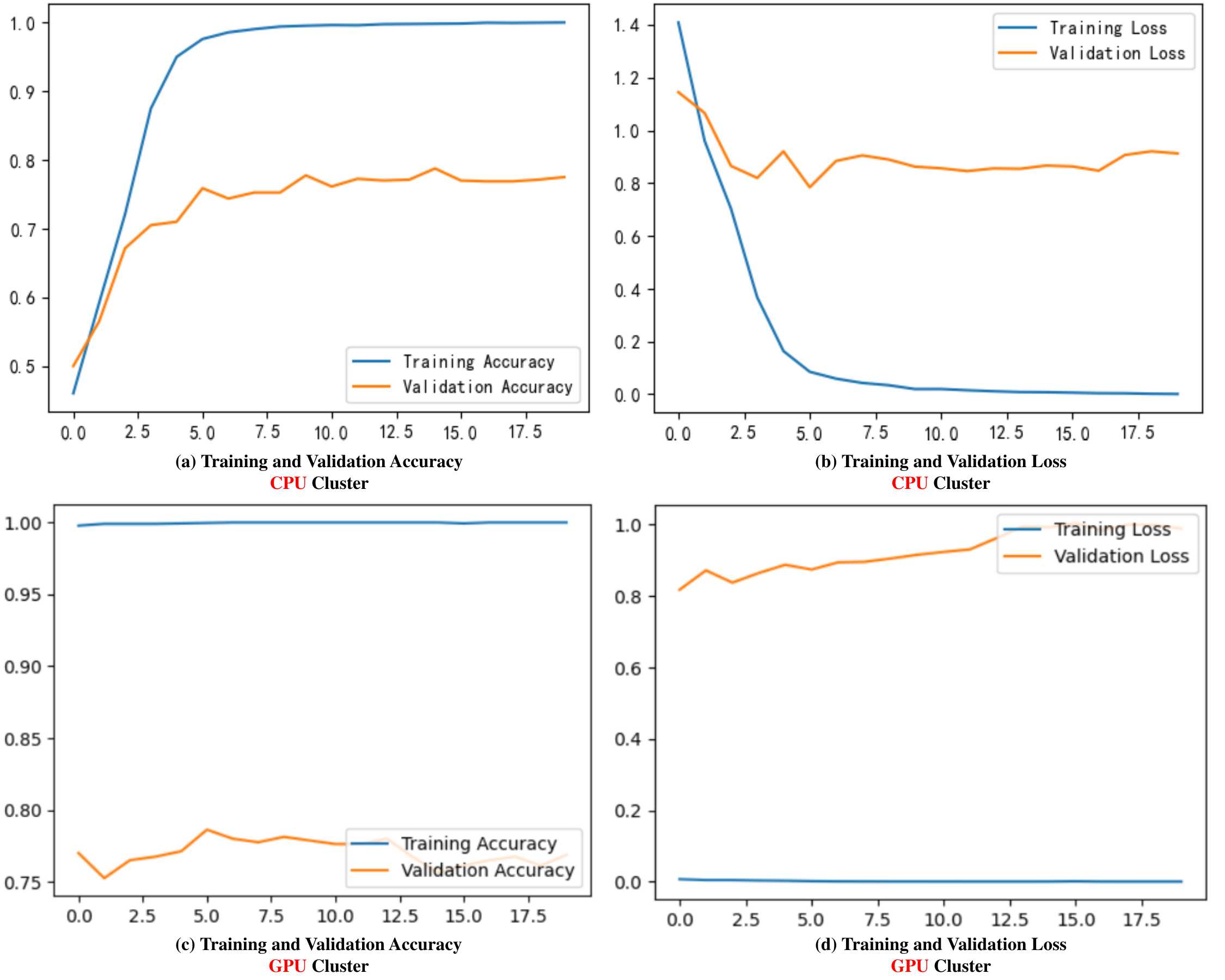}

 \caption{\textit{Training and validation accuracy and loss for the same task running by the CPU and GPU cluster respectively. 
 }} 
\label{fig:modelCOmpare1}
 \vspace{-.1in}
\end{wrapfigure} 

For serverless computing, we use AWS Lambda \cite{AmazonLambda} and AWS API Gateway \cite{AmazonAPIGateway} for the \ac{ML} application serverless implementation. The applications are set up under \texttt{x86\_64} architecture with Python 3.9.1 environment. The inference function is coded as the Lambda function and a route \texttt{img\_classify} is defined as a trigger in the API gateway.
As AWS Lambda allows 128MB to 3GB for the provisioned memory, we set up two experimental environments with 2GB and 3GB memory. The timeout is set as 50 seconds which is the same as the local web server. The \ac{ML} application is transferred as a Docker container and the runtime of the TensorFlow Keras is defined inside the DockerFile.

For containers, we follow the Dockerfiles on NVIDIA official repository%
to set up a virtual environment in the existing Windows server. After the WSL2 is successfully installed, we can set up the Docker environment inside the Ubuntu instance. Second, we installed the NVIDIA Docker Toolkit from the official repository,%
so we can run ML tasks with Docker. Third, we further modify the Docker files to make the tasks run automatically. We generated a \texttt{requirement.txt} file from the Python script we used for model training. 
We modified the existing Dockerfile by adding the RUN \texttt{pip install --no-cache-dir -r requirements.txt} and COPY in the Python script to the target directory. Lastly, we defined the starting command to run the Python script so that the Docker container can automatically run the task when the image is activated.

\textbf{ML model and dataset.} We use an image classification model, Xception \cite{chollet2017xception}, as the base model to test different implementation environments. Xceptioin can reach up to 95\% accuracy rate for image classification tasks and requires a minimum of 110.9MB for storage and the inference overhead is 109.4 ms. Xception model is suitable for our experiment since it demands intensive computational power in model development and its model size is also significantly larger than regular web applications. Therefore, we use it as the test base for our evaluation. The experiment uses an open-source training data set \cite{xceptionTrainData}, containing 299*299 4000 images and labeled as 1000 images for cats, 1000 images for chook, 1000 images for dogs, and 1000 images for horses. The data set is split 80\% (3200 out of 4000 images) into samples used to train the models and the rest (800 images) for validation.

\textbf{Metrics.} For model development, we focus on \textit{accuracy and loss} of the trained model. For model deployment, we care about \textit{session length, response time, and failed rate} when placing \ac{ML} application's requests in different computing platforms.

\subsection{Model Development}
\label{subsec:development}

We use Google TensorFlow \cite{abadi2016tensorflow} to perform the training of the model. 
A 20-fold cross-validation technique is used to evaluate the models during training. The training dataset is divided into 20 subsets: 16 are used for training and the remaining 4 are used for testing. This process is repeated until all samples of the training dataset have been used. The accuracy of the model is the average accuracy observed in each iteration.
The same training task is performed on the same cluster with CPU and GPU respectively.

\textbf{Accuracy and loss.} 
Figure \ref{fig:modelCOmpare1} compares the CPU and GPU clusters' accuracy and loss. Apparently, after 20 epochs, both models reach 99\% training accuracy and 70\% - 80\% validation accuracy and around 0.9 validation Loss. CPU and GPU clusters can finish the training tasks, however, from chart \textbf{(a)} and \textbf{(b)}, we see that for the CPU cluster, it took 9-10 epochs to get the final stable accuracy and Loss. On the other hand, as shown in \textbf{(c)} and \textbf{(d)}, it only takes 1-2 epochs to reach a final and stable accuracy. Even though it took around 10 epochs to get the final Loss, the deviation for the Loss is obviously smaller than the CPU cluster.

\subsection{Model Deployment}
\label{subsec:deployment}

\begin{wrapfigure}{r}{0.8\linewidth}
  
  \centering  
  \captionsetup[subfigure]{width=4cm}
  \subfloat[The failure rate on the local web server and in-house data center]
  {\includegraphics[width=0.2\textwidth]{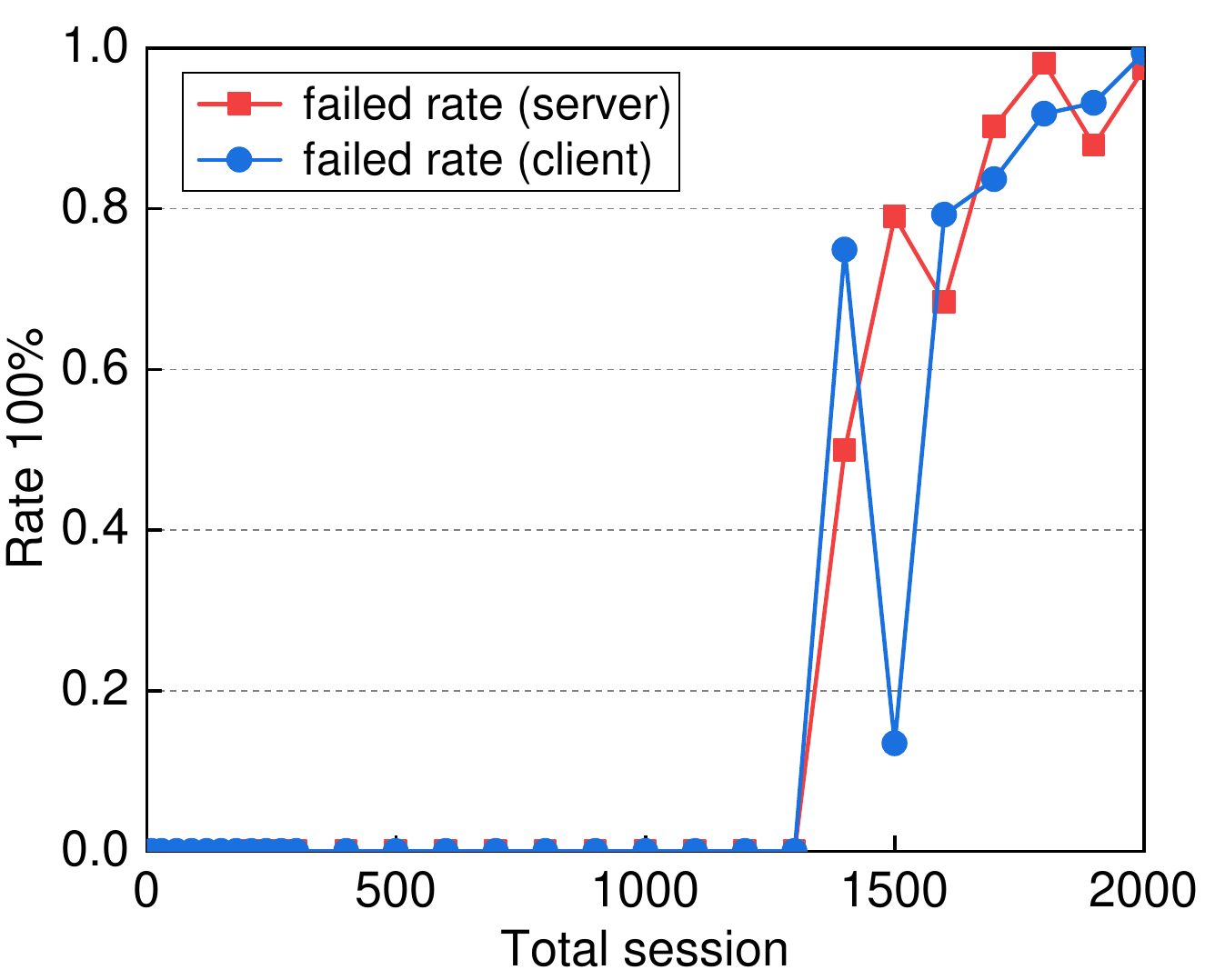}
  \label{fig: fail rate}}
 \hfill
   \hspace{-0.1in}
   \subfloat[The session length on local web server]{\includegraphics[width=0.2\textwidth]{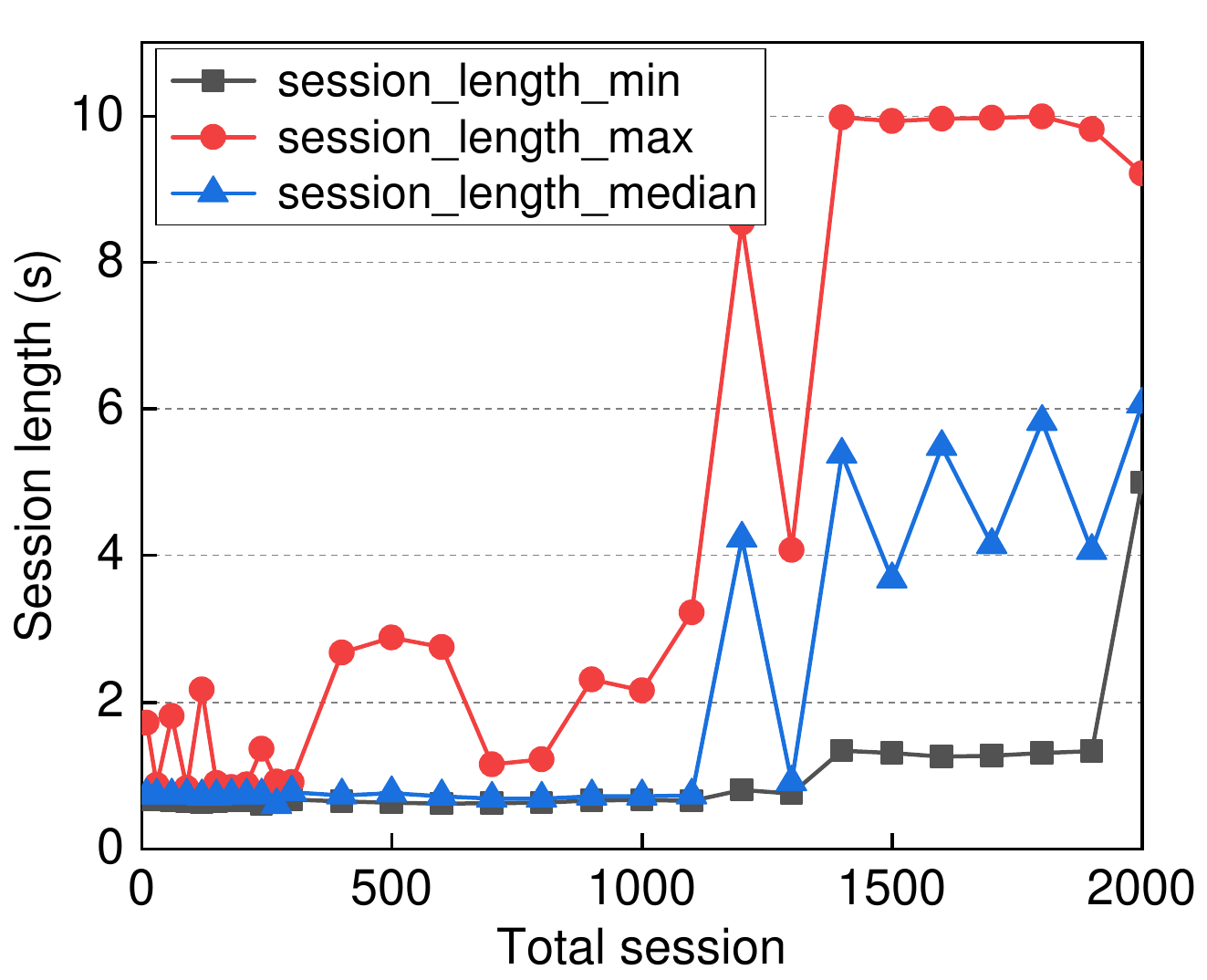}
  \label{fig: session length web server}}
 \hfill
     \hspace{-0.1in}
     \subfloat[The session length on in-house data center]{\includegraphics[width=0.2\textwidth]{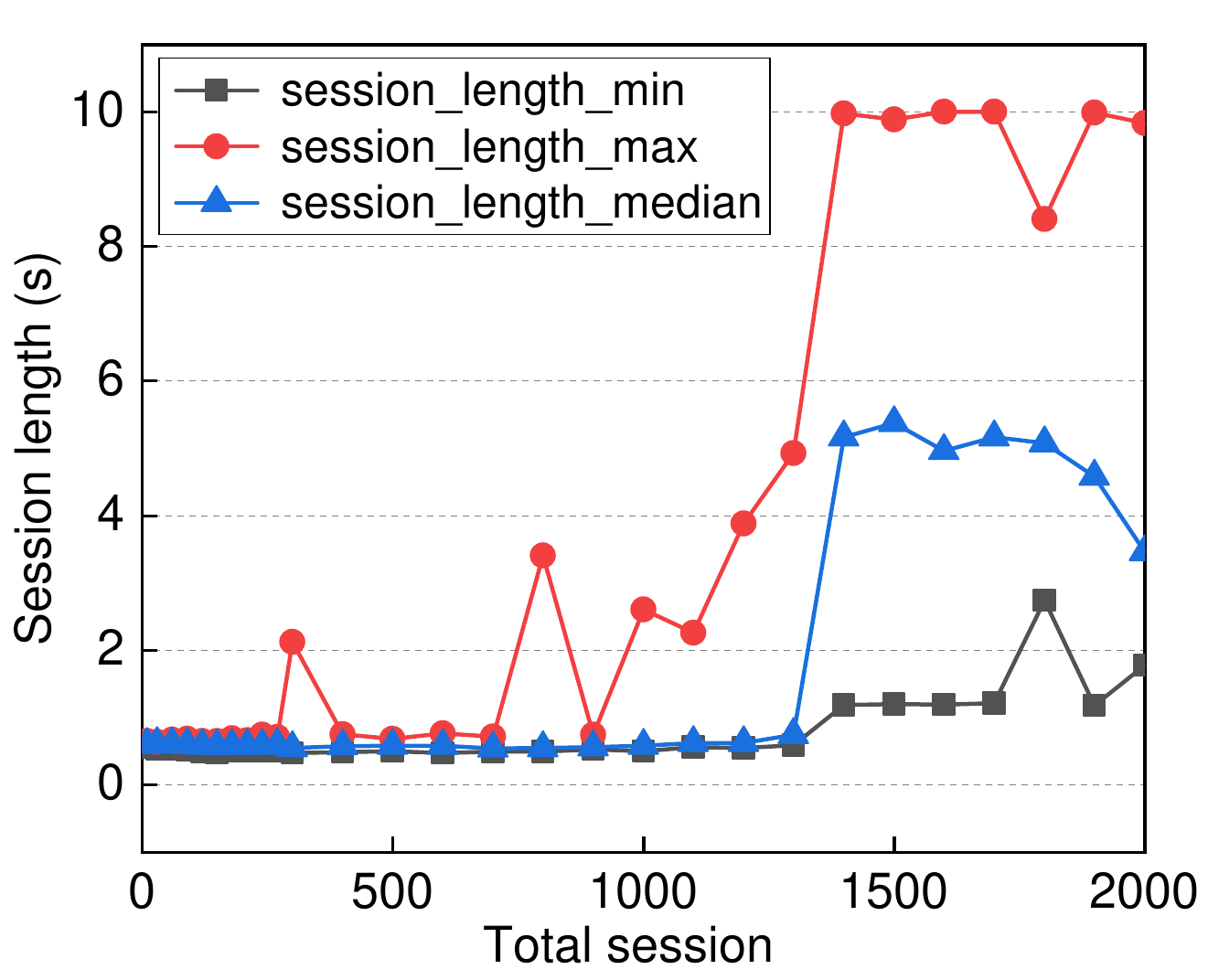}
   \label{fig: session length data center}}
   \vspace{-0.1in}
  \caption{\textit{The performance of a Flask API server implemented on a local web server and in-house data center. \textbf{(a)} shows the failure rate of the local web server and the in-house data center. \textbf{(b)} and \textbf{(c)} show the session length of the local web server and the in-house data center, respectively.
  Total sessions start from  10 requests per 180 seconds to 2,000 requests per 180 seconds against the Xception model.}}
  \label{fig: flask api web server and data center}
 \vspace{-0.15in}  
\end{wrapfigure}

We set up a real-world hybrid infrastructure for the seamless implementation of \ac{ML} applications among local servers, data centers, and AWS Lambda serverless environment. 
To generate the request load, we use a cloud-based performance scripting client, Artillery \cite{artillery}. 
We implement \ac{ML} application of Xception on different platforms and measure the performance with different request frequencies and request data sizes. We increase the load from 10 to 7000 within 180 seconds.

\textbf{RESTful service.}

Figure \ref{fig: flask api web server and data center} shows the load test results of Flask API implemented on the local web server and the in-house data center. From Figure \ref{fig: fail rate} we see that the local web server and in-house data center achieve a similar failed rate when total sessions reach 1300. From Figures \ref{fig: session length web server} and \ref{fig: session length data center} we can see that the session length in the local web server surges when total sessions reach 1200, whereas the session length in the in-house data center increases when total sessions reach 1400. This is because the web server implements a single thread to run the \ac{ML} application.

\textbf{Serverless computing.} 

\begin{wrapfigure}{r}{0.8\linewidth}
    \vspace{-0.2in}
  \centering 
   \captionsetup[subfigure]{width=4cm}
  \subfloat[The failed rate changes on AWS Lambda 2GB and 3GB respectively]
  {\includegraphics[width=0.21\textwidth]{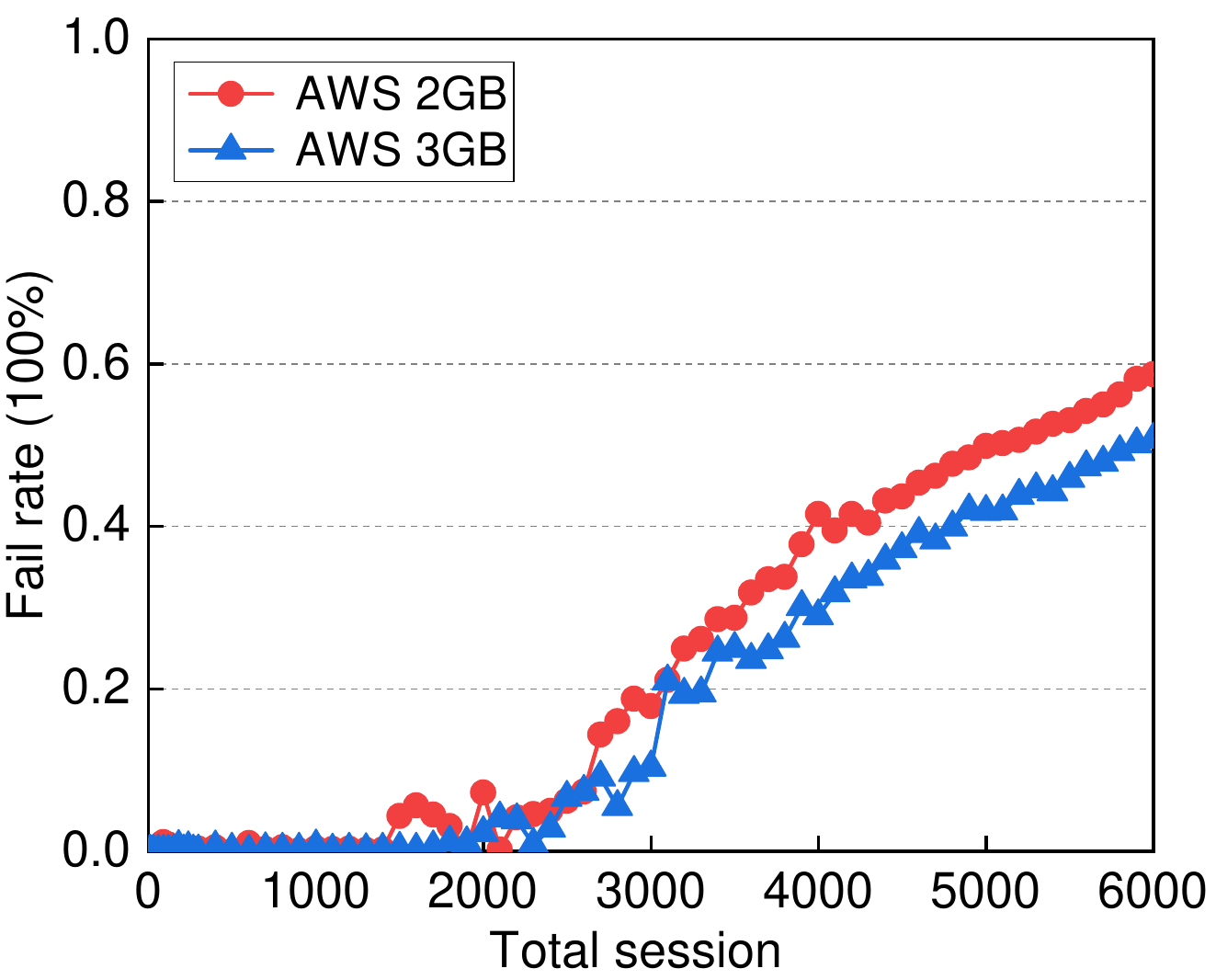}}
    \hfill
 \hspace{-0.1in}
   \subfloat[Session length and response time for AWS 2GB Memory Provisioned]{\includegraphics[width=0.21\textwidth]{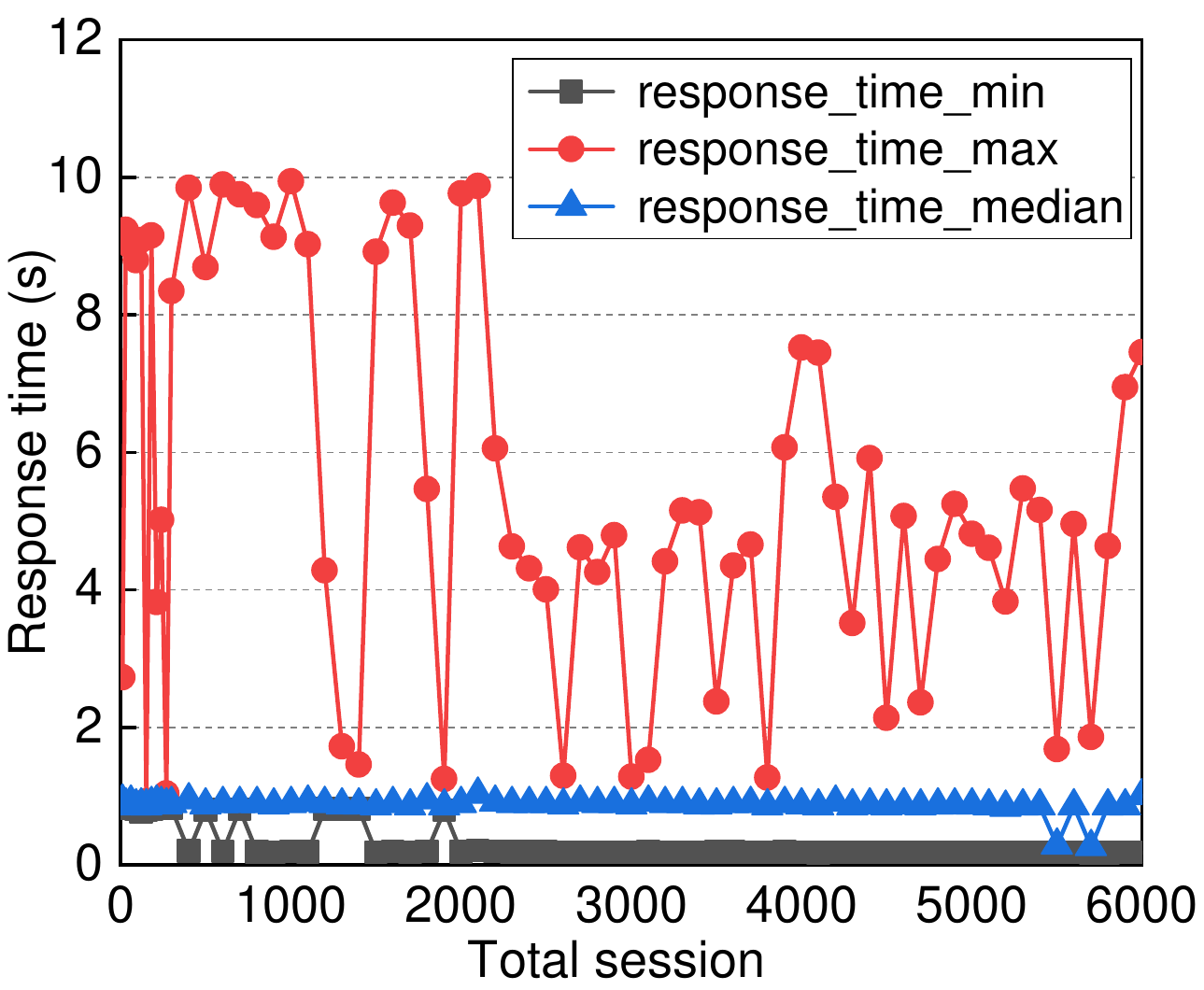}
   \label{fig:scheduling latency}}
   \hfill
     \subfloat[Session length and response time for AWS 3GB Memory Provisioned]{\includegraphics[width=0.21\textwidth]{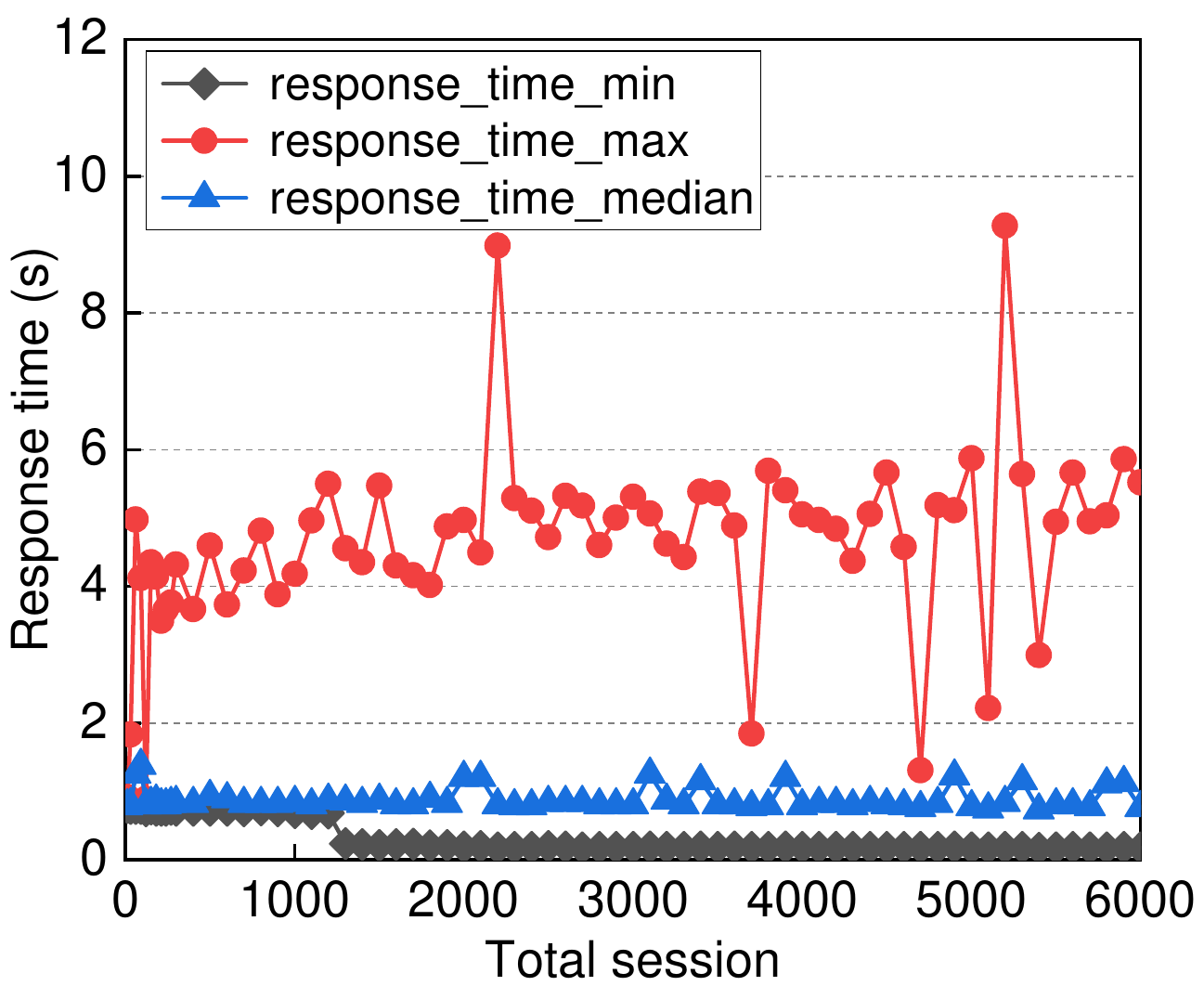}
   \label{fig:invocation_time_1}}
   \vspace{-0.1in}
  \label{fig:serverlessDocker}
  \caption{\textit{The results show the comparison for the same Xception application Docker implemented on AWS Lambda with 2GB and 3GB provisioned respectively. \textbf{(a)} shows the failed rate directly impacted by the provisioned memory. \textbf{(b)} and \textbf{(c)} show that the application latency is not affected much by the load changes.}}
 \vspace{-0.15in}
  \label{fig: serverless}
\end{wrapfigure}

Figure \ref{fig: serverless} shows the results of session length, and response time of AWS Lambda stack implemented with 2GB and 3GB Docker. The performance has improved significantly on the AWS Lambda stack. We can see that the median response time stays around 300-500 ms even when the request frequency increases up to 6000 per 180 seconds and the failed rate reaches up to 60\% when the request frequency rises to 6000 per 180 seconds. The results on AWS Lambda outperform the traditional RESTful API implementations.

\textbf{Hybrid Infrastructure.}

\begin{wrapfigure}{r}{0.8\linewidth}
    \vspace{-0.2in}
    \centering
    \begin{minipage}{0.25\textwidth}
        \centering
        \includegraphics[width=\textwidth]{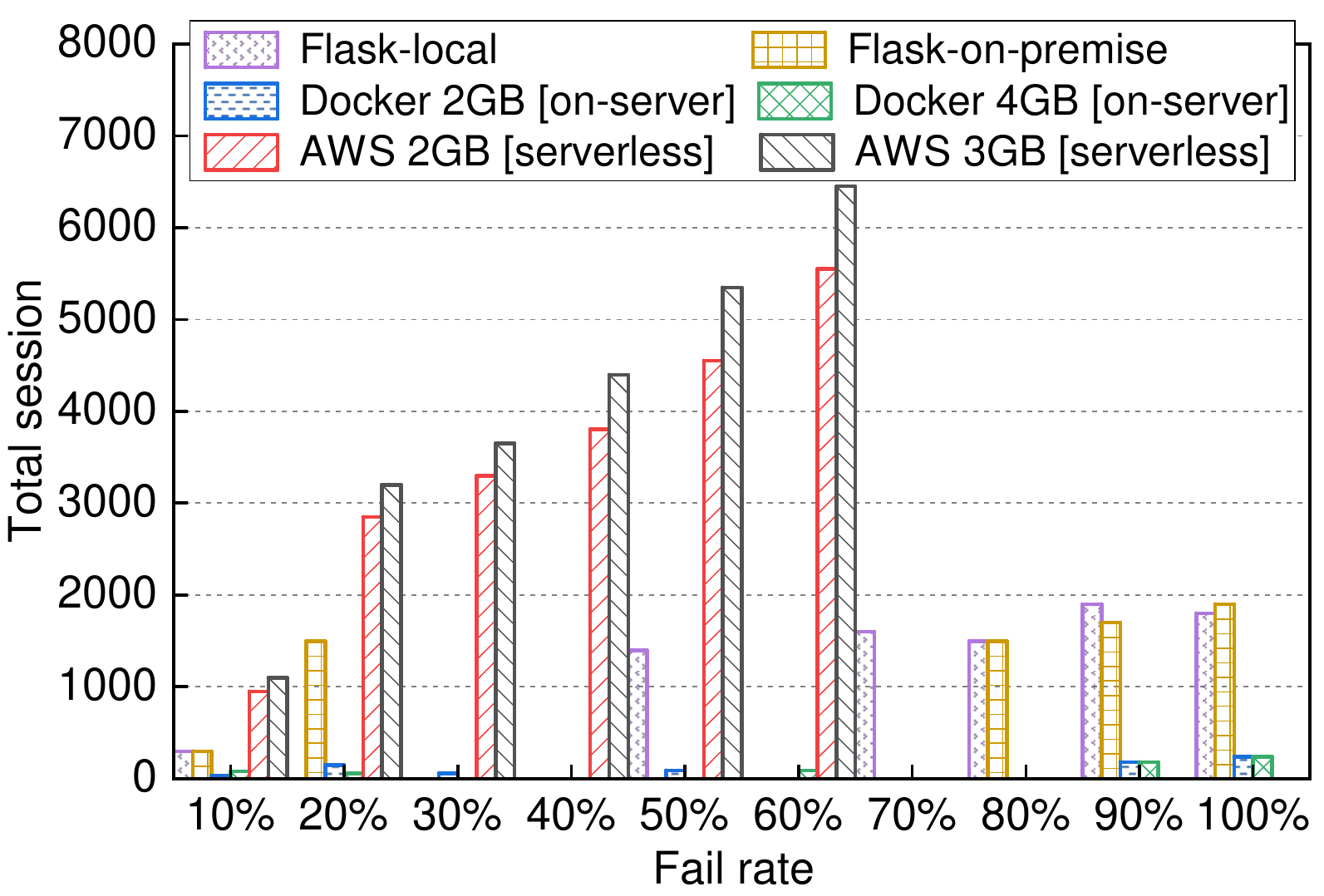}
        \vspace{-0.2in}
        \caption{\textit{Failed rates of different compute platforms when different numbers of sessions are created.}} 
        \label{fig:failed_rate_bar}
    \end{minipage}\hfill
    \begin{minipage}{0.25\textwidth}
        \centering
        \includegraphics[width=\textwidth]{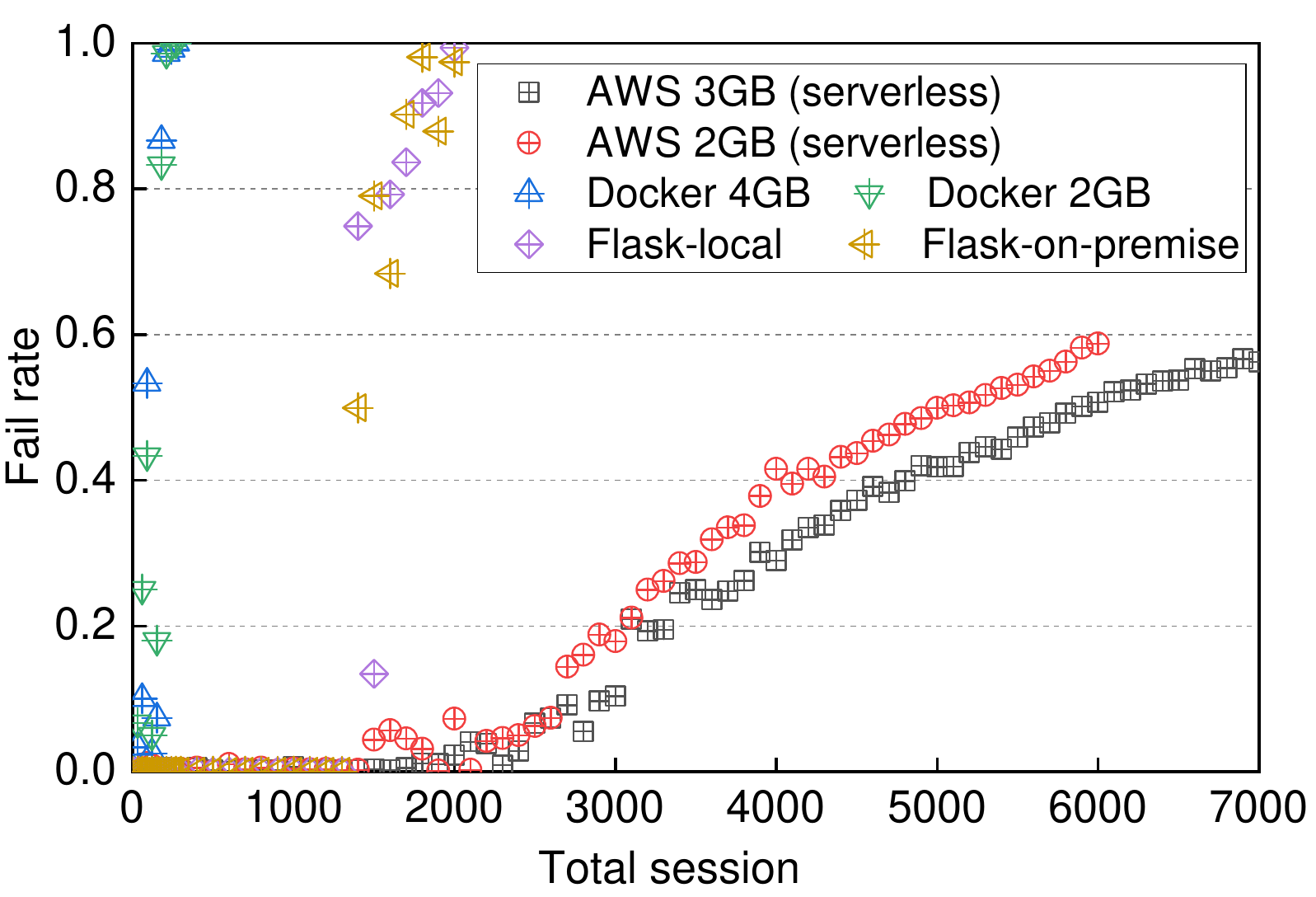}
        \vspace{-0.2in}
        \caption{\textit{Distributions of failed rates of different compute platforms when different numbers of sessions are created.}} 
        \label{fig:failed_rate_dot}
    \end{minipage}\hfill
    \begin{minipage}{0.25\textwidth}
        \vspace{.06in}\includegraphics[width=\textwidth]{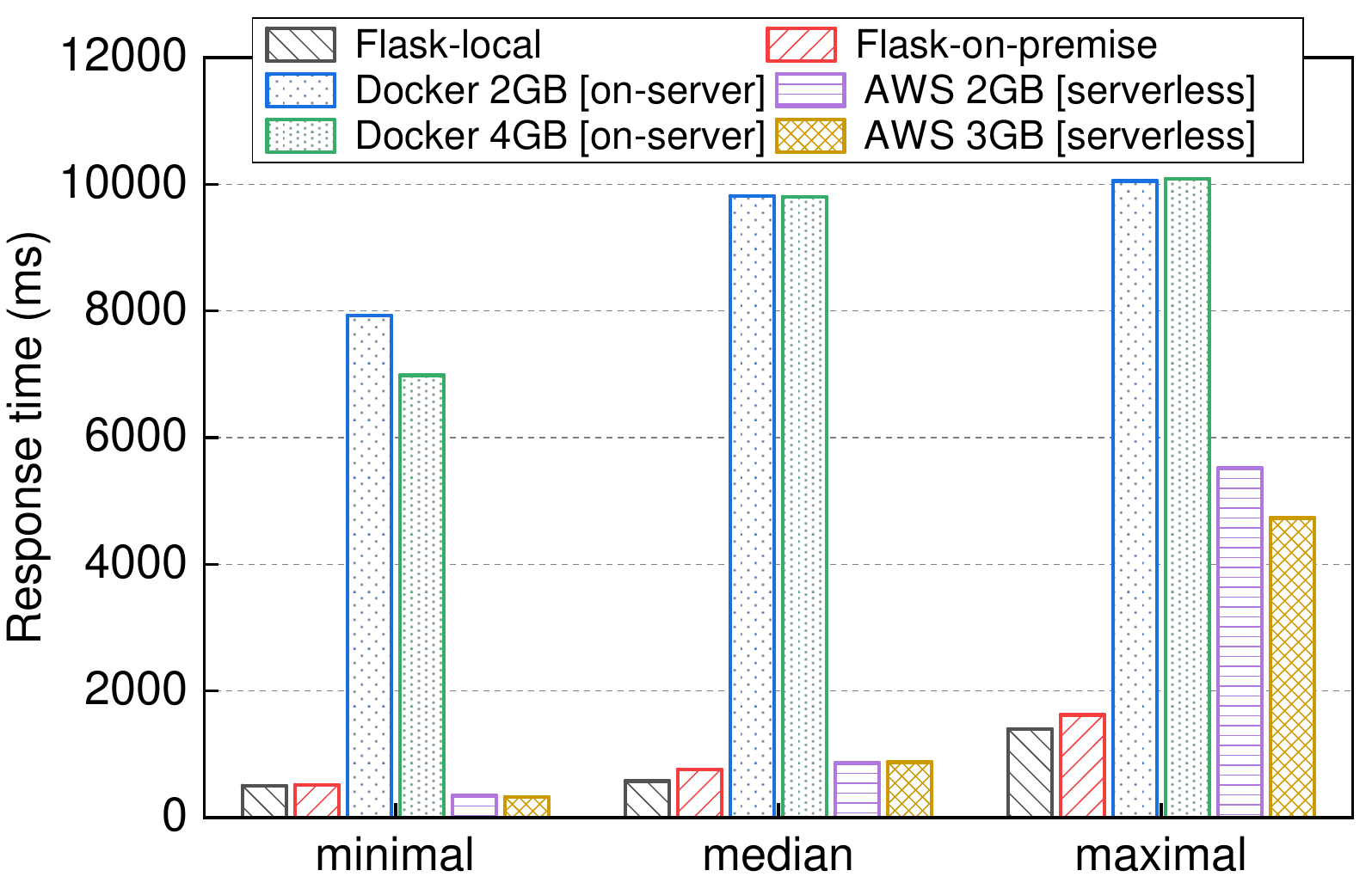}
        
        \caption{\textit{This figure shows that the Flask API outperforms the RESTful implementation in terms of response time.}}
        \label{fig:responseTimeCompare}
    \end{minipage}
     \vspace{-0.1in}
\end{wrapfigure}

Figure \ref{fig:failed_rate_bar} and Figure \ref{fig:failed_rate_dot} show the comparison of the failed rate of running \ac{ML} applications on different computing platforms. It is obvious that as the request frequency increases, serverless computing shows an acceptable failed rate. In addition, as the provisioned memory increased from 2GB to 3GB for serverless computing, the failed rate decreased. Therefore, serverless computing tends to be a good solution for high request frequencies. This is because serverless computing can offer unlimited resources for running \ac{ML} applications as well as an efficient orchestration method to trade-off the required computational resources to the provisioned memory.

Figure \ref{fig:responseTimeCompare} shows the comparison of the response time of running \ac{ML} applications on different computing platforms. Clearly, the Flask API outperforms the rest. This is because the Flask API can offer local computation for \ac{ML} applications to directly call the inference function. Both Docker and AWS serverless solutions cause longer response times because of the Docker container activation overhead.

\section{Conclusion and Future Work}
\label{sec:conclusion}

Traditional ML systems focus on one particular stage or phase of the life cycle of ML applications. These systems often aim at
optimizing model training or accelerating model inference, and
they frequently assume homogeneous infrastructure for model
training or model inference, which may not always reflect
real-world scenarios. In this paper, we present {\systemNameAbbr}, an end-to-end resource-aware scheduler that schedules the optimal resources (e.g., container, virtual machine, or serverless process) for different ML application requests in a hybrid infrastructure. {\systemNameAbbr} leverages Docker to containerize the stages in model development to offer plug-and-go provisioning. {\systemNameAbbr} proposes an empirical dynamic placing algorithm that intelligently places requests based on their unique characteristics (e.g., request frequencies and request data sizes).

We have conducted preliminary experiments to demonstrate StraightLine's effectiveness in reducing response time and failure rate. In the future, we plan to explore the following directions: (1) evaluate StraightLine across diverse scenarios and workloads and compare its performance with alternative approaches; (2) enhance StraightLine's empirical dynamic placing algorithm to consider additional parameters of the models and characteristics of the applications (e.g., SLOs); and (3) further refine StraightLine to dynamically allocate resources based on real-time demand fluctuations and workload patterns, ensuring optimal resource utilization and performance.




\bibliographystyle{IEEEtran}
\bibliography{References}

\end{document}